\documentclass[aps, prd, twocolumn, superscriptaddress]{revtex4}
\usepackage{graphicx}
\usepackage{dcolumn}
\usepackage{amssymb}

\begin{document}

\newcommand{\fs}{f_{10}}
\newcommand{\As}{A_{\mathrm{s}}}
\newcommand{\At}{A_{\mathrm{t}}}
\newcommand{\ns}{n_{\mathrm{s}}}
\newcommand{\nt}{n_{\mathrm{t}}}
\newcommand{\Obhh}{\Omega_{\mathrm{b}}h^{2}}
\newcommand{\Omhh}{\Omega_{\mathrm{m}}h^{2}}
\newcommand{\Ol}{\Omega_{\Lambda}}
\newcommand{\VEV}{\phi_{0}}
\newcommand{\conj}{^*}
\newcommand{\vect}[1]{\mathbf{#1}}
\newcommand{\half}{\frac{1}{2}}
\newcommand{\FT}[1]{\tilde{#1}}
\newcommand{\tauOS}{\tau_{\xi=0}}
\newcommand{\Rmax}{R_{\mathrm{max}}}
\newcommand{\CMB}{\textsc{cmb}}
\newcommand{\CMBEASY}{\textsc{cmbeasy}}
\newcommand{\CAMB}{\textsc{camb}}
\newcommand{\TT}{\textsc{tt}}
\newcommand{\TE}{\textsc{te}}
\newcommand{\EE}{\textsc{ee}}
\newcommand{\BB}{\textsc{bb}}
\newcommand{\B}{\textsc{b}}
\newcommand{\E}{\textsc{e}}
\newcommand{\BOOMERANG}{\textsc{boomerang}}
\newcommand{\CLOVER}{\textsc{C}$\ell$\textsc{over}}
\newcommand{\DASI}{\textsc{dasi}}
\newcommand{\EBEX}{\textsc{ebex}}
\newcommand{\POLARBEAR}{\textsc{p}olar\textsc{b}ea\textsc{r}}
\newcommand{\QUAD}{\textsc{qu}a\textsc{d}}
\newcommand{\QUIET}{\textsc{quiet}}
\newcommand{\SPIDER}{\textsc{spider}}
\newcommand{\WMAP}{\textsc{wmap}}
\newcommand{\GUT}{\textsc{gut}}
\newcommand{\CPU}{\textsc{cpu}}
\newcommand{\UETC}{\textsc{uetc}}

\title{CMB polarization power spectra contributions from a network of cosmic strings}

\newcommand{\addressSussex}{Department of Physics \&
Astronomy, University of Sussex, Brighton, BN1 9QH, United Kingdom}

\author{Neil Bevis} 
\email{n.a.bevis@sussex.ac.uk}
\affiliation{\addressSussex}

\author{Mark Hindmarsh} 
\email{m.b.hindmarsh@sussex.ac.uk}
\affiliation{\addressSussex}

\author{Martin Kunz}
\email{martin.kunz@physics.unige.ch}
\affiliation{D\'epartement de Physique Th\'eorique, Universit\'e de Gen\`eve, 1211
Gen\`eve 4, Switzerland}

\author{Jon Urrestilla}
\email{j.urrestilla@sussex.ac.uk}
\affiliation{\addressSussex}

\date{11 August 2007}

\begin{abstract}
We present the first calculation of the possible (local) cosmic string contribution to the cosmic microwave background polarization spectra from simulations of a string network (rather than a stochastic collection of unconnected string segments). We use field theory simulations of the Abelian Higgs model to represent local U(1) strings, including their radiative decay and microphysics. Relative to previous estimates, our calculations show a shift in power to larger angular scales, making the chance of a future cosmic string detection from the \B-mode polarization slightly greater. We explore a future ground-based polarization detector, taking the \CLOVER\ project as our example. In the null hypothesis (that cosmic strings make a zero contribution) we find that \CLOVER\ should limit the string tension $\mu$ to $G\mu<0.12\times10^{-6}$ (where $G$ is the gravitational constant), above which it is likely that a detection would be possible.
\end{abstract}

\maketitle


\section{Introduction}

There is an increasingly strong observational case for the inflationary paradigm, including precise measurements of the cosmic microwave background (\CMB) radiation. Many viable inflation models have been constructed, under a wide range of physical theories, but a prediction in many interesting cases is that a network of cosmic strings \cite{Vilenkin:1994book,Hindmarsh:1994re} should exist after inflation. This includes all viable inflation models under grand unified theories (\GUT s) below a certain complexity \cite{Jeannerot:2003qv} as well as brane inflation models in string theory \cite{Copeland:2003bj,Sarangi:2002yt,Jones:2003da,Dvali:2003zj}. However, the predicted cosmic strings would interact with radiation and matter via gravity and so source variations in the temperature of the \CMB\ in addition to those seeded directly by inflation. If their energy per unit length is too large then they destroy the fit to data yielded by the inflationary component, but current measurements still allow the string component to the temperature power spectrum to be about $10\%$ (dependent upon angular scale) \cite{Bevis:2007gh, Wyman:2005tu, Battye:2006pk}. Hence, while strings may be important for future \CMB\ measurements, they produce at most a secondary component of the temperature anisotropies, which is shrouded by that of inflation and is therefore difficult to detect. 

However, the polarization of the \CMB\ radiation contains information in addition to that from the temperature variations and it has been shown that inflation contributes only weakly to angular polarization patterns of a so-called magnetic nature \cite{Kamionkowski:1996ks}. \CMB\ polarization is hence a particularly interesting means to detect anisotropy components that are sub-dominant in the temperature power spectrum, since they may still give the primary contribution to polarization patterns of this type. Two important examples of such components are primordial gravitational waves and the case of primary interest here: cosmic strings \cite{Seljak:2006hi}. In this paper we present the \CMB\ polarization power spectra contributions from the latter, calculated for the first time using classical field theory simulations of a cosmic string network. The method for our calculations is discussed in more detail in \cite{Bevis:2006mj}, in which we limited the results presented to the temperature anisotropy. We reveal our polarization results in this separate publication because our discussion relies on results from \cite{Bevis:2007gh} but also because the nature of previous temperature and polarization calculations are different and these are not only the first polarization results to stem from field simulations of cosmic strings, they are the first to use actual simulations of a cosmic string \emph{network}. Finally we also wish to highlight that polarization offers exciting future observational possibilities and deserves a detailed discussion.


\section{CMB Calculation method}
\label{sec:method}

\subsection{Brief overview}

For this work we make use of the fact that the measured \CMB\ anisotropies are only of order $1$ part in $10^{5}$ and therefore that the perturbations in the radiation and matter do not have a noticeable effect on the string dynamics. Simulations of cosmic strings can therefore be performed in a homogeneous Friedman-Robertson-Walker universe with negligible loss of accuracy. The energy-momentum tensor of the cosmic strings from these simulations is then used to source metric perturbations via the Einstein equations and so creates perturbations in the radiation and matter. The low level of the perturbations also implies that any interaction between these string-induced perturbations and the primordially seeded inflationary ones can be ignored, and therefore that the cosmic string contribution to the \CMB\ can be calculated separately. With negligible coupling between them, the two contributions are statistically independent and the total power spectrum is simply the sum of the power spectra from each of the two sources. 

The contribution from inflation has been well studied and a number of codes exist which accurately solve the Einstein-Boltzmann equations and, using a modern desktop computer, results for both temperature and polarization power spectra can be obtained in a few seconds. The cosmic string contribution is more challenging due to the non-linear string dynamics and the enormous range of scales involved in the problem. There is strong evidence that cosmic string networks enter an attractor solution or scaling regime \cite{Kibble:1984hp, Albrecht:1989mk, Bennett:1989yp, Allen:1990tv, Vincent:1997cx, Moore:2001px, Martins:2005es, Ringeval:2005kr, Bevis:2006mj} in which the average string separation is comparable to and scales with the causal horizon size. However, the width of strings corresponds inversely to their energy scale, which may be $\sim 10^{16}\;\textrm{GeV}$, and therefore the width is many orders of magnitude smaller than their separation at times of importance for \CMB\ calculations. The problem is hence not directly solvable with current or near future technology. Indeed, the dynamics of strings on scales much smaller than the horizon is not well understood \cite{Vincent:1996rb, Vincent:1997cx, Martins:2005es, Ringeval:2005kr}, with questions relating to the production of string loops and of small-scale structure on long strings. There is not even a consensus on whether gravitational radiation or high energy particle emission is the dominant means via which strings decay \cite{Vincent:1996rb, Vincent:1997cx, Moore:2001px}. These facts limit the reliability of gravitational radiation predictions but fortunately \CMB\ anisotropies are sourced mainly by the distribution of long strings on scales close to the horizon, which are more easily simulated.

The only previously published \CMB\ polarization results for (local) cosmic strings \cite{Wyman:2005tu} (and the derived \cite{Battye:1998xe,Seljak:2006hi}) use the computationally rapid method of Albrecht et al. \cite{Albrecht:1997nt},  which is reliant upon two levels of approximation. Firstly, the Nambu-Goto approximation is applied, such that the strings are taken to have infinitesimal width. Secondly, rather than simulating a string network, the strings are represented by a series of straight and unconnected string segments, moving with stochastic velocities. The sub-horizon decay of strings is modeled by the random removal of segments such that the total string length varies as it would for a network according to the above scaling law. Then the angle-averaged energy-momentum tensor of the segments is used to source the cosmological perturbations. The results for the temperature power spectrum obtained from this unconnected segment model are in broad agreement with those obtained from actual simulations of a connected string network, evolved using the Nambu-Goto equations of motion \cite{Contaldi:1998mx}. This gives confidence in the applicability of the computationally rapid method, however for the case of polarization, there are no published results from Nambu-Goto simulations with which to compare those found using the unconnected segment model.

However, the Nambu-Goto simulations themselves are not complete and questions have been raised over the accuracy with which they represent string loop production and the string decay \cite{Vincent:1996rb, Vincent:1997cx, Olum:1998ag, Bevis:2006mj}. In \cite{Bevis:2006mj} we presented the first calculations of the temperature power spectrum for (local) cosmic strings to stem from field-theoretic simulations. Those calculations included \CMB\ polarization, as is required for the accurate calculation of the temperature anisotropy and we hence present our polarization results here with minimal discussion of the method employed. 

\begin{figure}
\resizebox{\columnwidth}{!}{\includegraphics{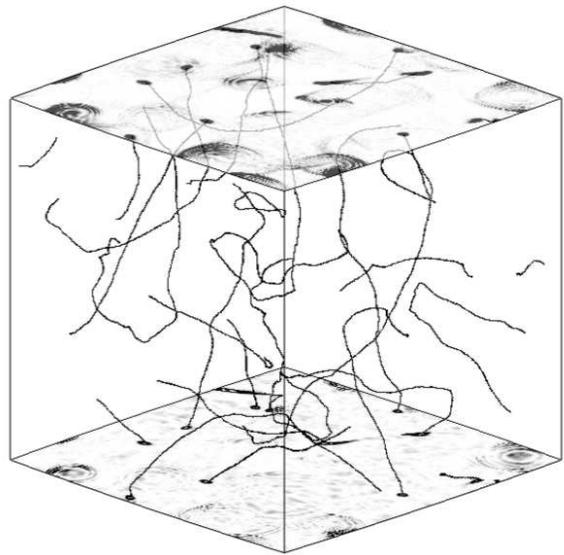}}
\caption{\label{fig:sims}A snapshot from an Abelian Higgs simulation in the matter era at a time when the horizon volume approximately fills the simulation box. The lines show the centers of the strings (found using the gauge-invariant phase-winding method of \cite{Kajantie:1998bg}) while the upper and lower surfaces highlight the additional presence of radiative decay, which must be included in an adhoc manner in Nambu-Goto simulations. The lower surface indicates regions of significant energy density due to the non-vacuum value of the Higgs field while the upper surface shows regions of significant energy from the quasi-magnetic field in the model (see \cite{Bevis:2006mj}). Note, however, that the strings themselves make the primary contribution to both of these types of energy and the contrast is chosen to highlight the radiation contribution. For example, the circular pattern seen on the left in these slices is due to the recent collapse of a string loop just above the bottom of the simulation (and is seen in both slices due to the periodic boundary conditions).}
\end{figure}

In short, by employing field simulations of the Abelian Higgs model, we have studied local U(1) strings with the string width resolved, accurately representing the associated microphysics, including a form of radiative decay, see Fig. \ref{fig:sims}. While this means that the simulations were limited to very small length scales at very early times, the scaling property enables statistical results to be translated to larger scales and later times and hence provide the information required for \CMB\ calculations. The corresponding extrapolation may be over many orders of magnitude, but it is justified by the behaviour of the strings in the simulations and it is not merely a theoretical assumption.  

We now present greater depth of the calculation procedure, in particular pointing out that scaling alone is not sufficient to enable current technology to perform the string simulations needed for \CMB\ calculations, but readers who are familiar with \cite{Bevis:2006mj}, or who are less concerned with these details, may wish to leave the following section for future reading.

\subsection{Greater detail}

Despite the use of the scaling property of cosmic strings, the simulations required for \CMB\ calculations could not be performed directly. The width of an Abelian Higgs string is a fixed physical scale and hence lessens rapidly relative to the physical horizon size, which varies as $\tau^{2}$ in the radiation era and $\tau^{3}$ in the matter era, where $\tau$ is the conformal time. Since scaling is broken at the radiation-matter transition, strings must be studied under both radiation and matter domination, but to do so over a sufficiently large range in $\tau$ is very challenging. This is a particular problem because strings will not form in the simulations until the horizon is much greater than their width and even then, the strings will take some time to reach the scaling regime. 

Our solution to this problem was to raise the coupling constants in the Abelian Higgs action to time dependent variables such that the string width $r$ grew slightly in physical (rather than comoving) units:
\begin{equation}
r_{\mathrm{phys.}} \propto a^{1-s},
\end{equation}
where $a$ is the cosmic scale factor and $s$ is a parameter that controls the growth. The true case has $s=1$ while a fixed comoving width is obtained by setting $s=0$. The later is perhaps the most straightforward to simulate and the authors of \cite{Moore:2001px} had previously studied string scaling using equations similar to those obtained via our method with $s=0$. The closest to the true case that proved possible using \mbox{$512\times512\times512$} simulation lattices in the matter era was $s=0.3$ \footnote{ Beyond $s=0.3$, a string network could not be formed and studied in the scaling regime for a sufficient range in $\tau$ for the required data to be extracted from the simulations} (with simulations performed using 64 processors of the UK National Cosmology Supercomputer \cite{cosmos-website}). However, the use of $s \neq 1$ causes a breach of the conservation law of the very energy-momentum tensor that sources the anisotropies, which may not have been an major problem for \cite{Moore:2001px} but is obviously a potential problem for \CMB\ calculations. Fortunately the effect can be investigated by simulating strings at a number of $s$ values, including the true $s=1$ case for the less difficult radiation era. Accordingly, in \cite{Bevis:2006mj} we showed that $s$ has only a minor effect on the string length behavior, the Fourier distribution of the energy-momentum tensor components and upon the \CMB\ results. The use of $s<1$ is therefore treated merely as a systematic source of uncertainty, comparable to the statistical uncertainties arising from realization-to-realization variations. Hence as a result of this procedure, the data required for accurate calculations of the \CMB\ power spectra were obtained from the simulations, with the results below being taken from simulations with $s=0.3$ in both radiation and matter eras.

If only power spectra are desired (rather than bi-spectra or maps), the unequal-time correlations of the energy momentum tensor $T_{\mu\nu}$:
\begin{equation}
\FT{U}_{\lambda\kappa\mu\nu}(\vect{k},\tau,\tau') 
= 
\left< \FT{T}_{\lambda\kappa}(\vect{k},\tau) \FT{T}_{\mu\nu}\conj(\vect{k},\tau') \right>,
\end{equation}
are all that is required to solve the Einstein-Boltzmann equations \cite{Pen:1997ae, Durrer:2001cg, Bevis:2006mj}. While there are $\half 10(10+1)=55$ such correlation functions, each complex functions of $3+1+1=5$ input variables, they are heavily constrained by scaling, statistical isotropy, causality and energy-momentum conservation. As a result they may be represented by merely 5 real functions of $2$ variables: $\FT{C}_{ab}(k\tau,k\tau')$ \cite{Pen:1997ae,Durrer:2001cg, Bevis:2006mj}. For example, statistical isotropy implies that only the magnitude of $\vect{k}$ is important for any expectation value, while scaling implies that the spatial distribution, on average, scales with the horizon and hence only the product $k\tau$ is important, rather than $k$ and $\tau$ individually \footnote{In fact $\tau\tau'$ remains important, setting the overall normalization of $\FT{U}$ and is used in the conversion between $\FT{U}$ and $\FT{C}$}. Of the 5 scaling function, 3 represent scalar degrees of freedom with $ab$ equal to 11, 12, 22 (with the 21 case given by $\tau \leftrightarrow \tau'$), while vector and tensor degrees of freedom give one function each. These unequal-time scaling functions (calculated under both matter and radiation domination) encode the information required to accurately calculate the power spectra contributions as a result of the sourcing $T_{\mu\nu}$. 

However, in order to solve the Einstein-Boltzmann equations it is convenient to re-express these scaling functions, which like $U_{\lambda\kappa\mu\nu}$ are quadratic in $T_{\mu\nu}$, as linear functions $\FT{c}_{n}(k\tau)$ \cite{Turok:1996ud, Bevis:2006mj}:
\begin{equation}
 \FT{C}(k\tau,k\tau')
 = \sum_{n} \lambda_{n} 
    \FT{c}_{n}(k\tau) \;
    \FT{c}_{n}(k\tau').
\end{equation}
In the numerical case, $\FT{C}$ is known only at discrete values of $k\tau$ and $k\tau'$ and the above equation corresponds to a decomposition of the real, symmetric matrix $\FT{C}$ into its eigenvectors $c_{n}$ and eigenvalues $\lambda_{n}$. The eigenvectors then represent linear combinations of $T_{\mu\nu}$, which source metric perturbations via the Einstein equations. These were fed into a modified version of \CMBEASY\ \cite{Doran:2003sy}, and the \CMB\ power spectrum contribution from a particular eigenvector was found, with the total string contribution being the sum over these. Modifications to \CMBEASY\ included the additional presence of the source terms and also cosmological perturbations of a vector nature, which decay in the standard inflation case but are continuously sourced by cosmic strings. These were evolved using gauge-invariant perturbation equations obtained via the Hu and White total angular momentum method \cite{Hu:1997hp}. 


\section{Results}

\begin{figure}
\resizebox{\columnwidth}{!}{\includegraphics{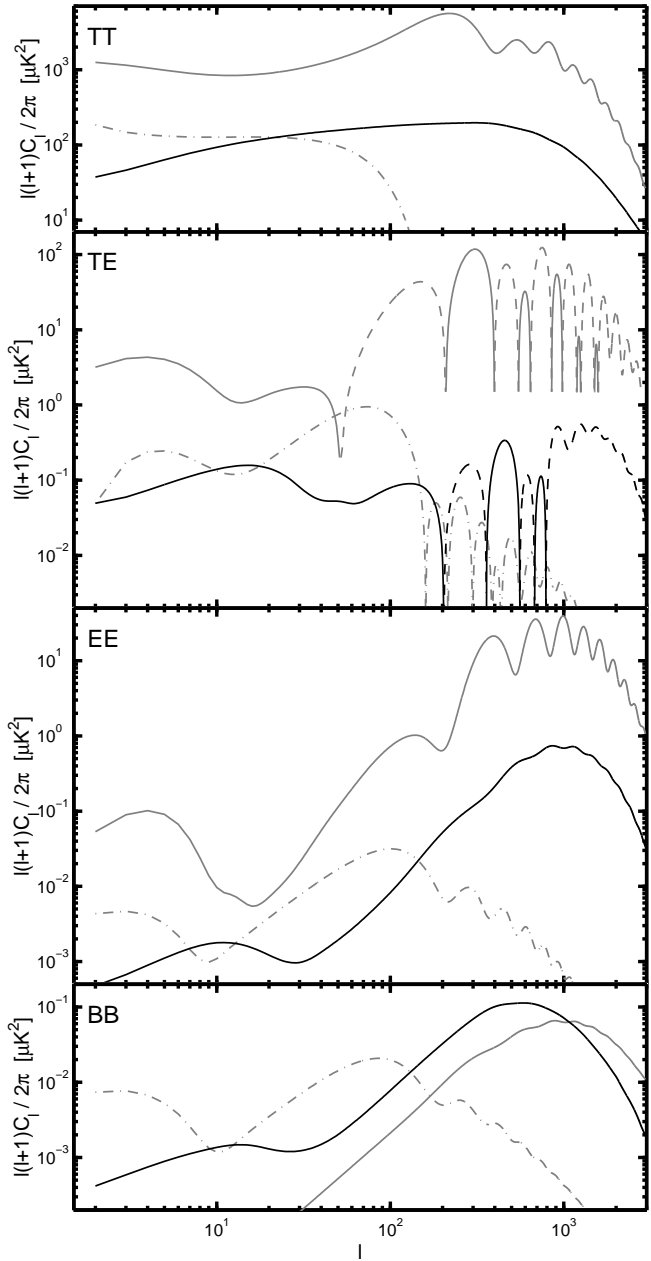}}
\caption{\label{fig:TEB}The \CMB temperature and polarization power spectra contributions from cosmic strings (black), inflationary scalar modes (gray, solid) and inflationary tensor modes (gray, dot-dashed). For the case of the \TE\ cross correlation, positive correlations are shown as solid lines and anti-correlations are shown as dashed lines, except that is for the inflationary tensor component for which the sign is not indicated here.}
\end{figure}

The results for both temperature and polarization power spectra are shown in Fig. \ref{fig:TEB}, and compared to the inflationary contribution, calculated here using \CAMB\ \cite{Lewis:1999bs}. The normalizations of the three quantities shown in the figure: the string contribution, the inflationary scalar contribution, and the inflationary tensor mode; are each free parameters. The inflationary tensor (or gravity wave) contribution has yet to be discovered, with the scalar mode alone giving an excellent fit to current data. Hence in the four plots shown, the inflationary scalar component is shown with its normalization $\As^{2}$ and other cosmological parameters chosen to match current \CMB\ data \cite{Kuo:2002ua, Jones:2005yb, Readhead:2004xg, Grainge:2002da, Hinshaw:2006ia}, without including strings or inflationary tensor modes (see endnote \footnote{Parameters obtained from a fit to \CMB\ data are: $h=0.74\pm3$, $\Obhh=0.0224\pm0.0007$, $\Omhh=0.127\pm0.007$, $\tau=0.09\pm0.03$, $\As^{2}=(22.0\pm1.2)\times 10^{-10}$ and $\ns=0.95\pm0.02$, with zero-running of $\ns$ and spatial-flatness assumed.} for parameters). 

The normalization of the string contribution is set at the $95\%$ upper bound allowed by current \CMB\ data, as calculated by the current authors in \cite{Bevis:2007gh}, using the above \CMB\ data together with parameter estimates from non-\CMB\ experiments: $h=0.72\pm0.08$ \cite{Freedman:2000cf} and $\Obhh=0.0214\pm0.0020$ \cite{Kirkman:2003uv}. This bound assumes negligible primordial tensor modes (although it is not especially sensitive to them) and corresponds to an $11\%$ contribution to the temperature (\TT) power spectrum at $\ell=10$ or fractional contribution $\fs=0.11$. For this string component, the normalization is related to the energy-scale of the associated theory, which determines the string tension $\mu$. It is conventional to express this in terms of $G\mu$ (where $G$ is the gravitational constant) with the power spectra proportional to $(G\mu)^{2}$. The results plotted here correspond to a value $G\mu = 0.7 \times 10^{-6}$ \cite{Bevis:2007gh} (and also $h=0.72$, $\Obhh=0.0214$, $\Ol=0.75$ \cite{Knop:2003iy}, spatial flatness and $\tau=0.1$). 

In a similar manner, the inflationary tensor mode normalization is at the $95\%$ upper bound set by the same data and method that we employed in \cite{Bevis:2007gh} to constrain the string component, without including strings while calculating the tensor case. This corresponds to an inflationary tensor contribution to the temperature (\TT) power spectrum at multipole $\ell=10$ of $15\%$ or a ratio of tensor-to-scalar primordial perturbation power spectra of $r=\At^{2}/\As^{2}=0.36$ (at comoving wavevector $k_{0}=0.01$ Mpc$^{-1}$). In this determination we used the single-field consistency equation \cite{Leach:2002ar} to give the tensor spectral tilt: $\nt = -r/8$, which implies $\nt=-0.045$ for the case plotted.

When interpreting the results, it is useful to consider the mechanism involved in creating \CMB\ polarization: Thomson scattering in the presence of a quadrupole intensity anisotropy, as seen by the scattering particle. This results in differing contributions being given to perpendicular polarization directions and therefore partially polarized emission. However, in the presence of very strong Thomson scattering, a quadrupole moment cannot be set up and hence there is no contribution until the universe begins to deionize. But as this occurs the frequency of scattering decreases, soon becoming negligible. Hence \CMB\ polarization stems primarily from a short period around deionization and therefore the polarization spectra are dominated by small angular scales. However, the radiation from stars results in partial reionization for more recent times, giving a small peak at large angular scales. These two features are clearly seen in all polarization graphs, for both strings and inflation, although the inflationary tensor modes do not contribute to very small scales.

\subsection{TT spectrum}

Before discussing the individual polarization spectra, the \TT\ power spectrum component from strings deserves a brief comment here. The string contribution is a broad peak between $\ell \approx 40$ and $\ell \approx 600$ (75\% of peak value), without the acoustic oscilations that are seen in the data and the matching inflationary scalar mode. In the inflationary case there are frozen super-horizon perturbations, which for a particular length scale enter the growing horizon and begin to oscillate at the same time. This gives a high degree of temporal coherence. In the string case, perturbations are \emph{continuously} sourced within the horizon by objects which themselves experience complex, non-linear dynamics. As a result no oscillations are seen in the plot for strings.

\subsection{EE spectrum}

Unlike the decomposition of a spin-1 vector field into curl-free and divergence-free parts, the decomposition of the spin-2 polarization field into so-called \E and \B modes is based upon the second angular derivatives of this 2D field. We first address the \E\ polarization power spectrum, which measures polarization patterns in which a minimum in the polarized intensity is accompanied by the greatest gradient being in a direction parallel or perpendicular to the plane of polarization. Only then will we discuss the \B\ mode, for which the greatest gradient is at $45^\circ$ to the polarization plane.

The cosmic string component to the \EE\ power spectrum is, very approximately, of the same form as that from inflationary scalar modes, although with much smaller oscillations at small scales. Since strings source perturbations most strongly on scales that are well within the horizon, whereas inflation seeds super-horizon perturbations that oscillate after they enter the horizon, the smaller oscillations are not surprising --- nor is the additional shift to 
higher multipoles. Some oscillations are expected since the electric polarization is heavily dominated at small scales by scalar modes (which oscillate), but decoherence almost completely erases the oscillations from the \EE\ power spectrum.

The logarithmic vertical axes of these four plots share a common log-interval per unit length and therefore a given distance traversed upwards corresponds to an increase by the same factor on each plot. It is hence clear that the amplitude of the string EE component relative to that from inflationary scalar modes is comparable to that of the temperature power spectrum, but since temperature data is likely to be significantly more precise than polarization data for the foreseeable future, measurements of the \EE\ data is not likely to place direct constraints upon cosmic strings for some time.

\subsection{BB spectrum}

The situation is quite different, however, for the \BB polarization power spectrum. Inflationary models only create scalar and tensor modes, while 
cosmic strings also create relatively large vector modes. The primordial scalar mode contributes to \BB\ polarization mainly \footnote{The primordial scalar mode can give rise to very small vector and tensor contributions in non-linear perturbation theory and so create small \BB\ spectrum contributions \cite{Bartolo:2007dh}, but they will not be relevant unless cosmic strings have a very small contribution indeed, and the cosmic shear signal is successfully \emph{cleaned}.} through the partial conversion of the \EE\ spectrum to a \BB\ contribution, which is due to gravitational lensing by matter perturbations \footnote{In principle cosmic strings and the matter perturbations that they seed contribute to the lensing of the inflationary power spectra, and the string spectra are also lensed. As the string perturbations are sub-dominant we neglect the latter contributions here, but this is an approximation that should be tested.}. Therefore the contribution from cosmic strings dominates the \BB\ spectrum for $\ell=150-1000$, despite them being constrained to be sub-dominant in the temperature power spectrum. At the current upper limit on the string normalization, the \BB\ contribution is roughly 3 times larger than the inflationary contribution in the range $\ell=150-500$ and despite it being likely that any real string contribution may have a lower normalization, a quite large reduction is required to stop them from dominating these scales. 

At low multipoles the reionization peak yields an even greater dominance over the lensing signal,  however, the inflationary tensor modes may contribute significantly to these scales. Indeed, if strings contribute at the $95\%$ upper bound quoted above and make up $11\%$ of the temperature power spectrum at $\ell=10$, while inflationary tensor modes contribute $15\%$ (although strictly, these two upper bounds should not be employed simultaneously), then the tensor component would be between 2 and 8 times larger over the range $20 \leq \ell \leq 100$, but about equal to that from strings for $\ell \approx 10$. As a result, if strings contributed at the level plotted but $r<0.05$ (as would be the case for supersymmetric hybrid inflation models \cite{Battye:2006pk}) then strings would dominate at all $\ell < 1000$.

\subsection{TE spectrum}

The \TE\ cross-correlation spectrum (between the temperature and electric polarization mode) shows significant oscillations in the string case, although decoherence suppresses them relative to the inflationary case. Both the inflation and string contributions are constant to within an order of magnitude for large scales and they both oscillate between anti-correlations and positive correlations for $100\lesssim \ell \lesssim 1000$. Note that the sign convention we chose for the \TE\ spectrum is that by the \WMAP\ team.

As is clear from the figure, the \TE\ spectrum contribution from cosmic strings is generally less significant relative to the inflationary contribution than in the \TT\ or \BB\ spectra. Although the oscillations seen for cosmic strings are $180^\circ$ out of phase with the those from the inflationary scalar mode, which may assist the use of \TE\ data in constraining cosmic strings, it is evident that the \TE\ spectrum not especially important for constraining the cosmic string component directly. Future \TE\ data will however, further constrain the inflationary scalar contribution to all spectra and reduce the cosmological parameter degeneracies which allows it to change and accommodate large cosmic string (or other topological defect) components while still fitting the data \cite{Bevis:2007gh, Bevis:2004wk}. 

\subsection{Uncertainty estimation}

Finally, it is of course the case that all these results incorporate, at some level, uncertainties relating to the random initial conditions used in the simulations, finite volume effects and the use of $s<1$ (see Sec. \ref{sec:method}). Our investigations into these and other possible uncertainties are described in the appendix.


\section{Comparison with previous local string results}

The only previously published calculations of the polarization power spectra contributions from local cosmic strings  
\cite{Wyman:2005tu} are based on the unconnected segment model. This model has a number of free parameters which were fixed by reference to Nambu-Goto simulations \cite{Allen:1990tv} and another phenomenological model, the velocity-dependent one-scale model \cite{Kibble:1984hp,Bennett:1985qt,Martins:1996jp}. In \cite{Bevis:2006mj} we pointed out that our simulations lead to temperature power spectrum results that are more biased to large scales than those of the unconnected segment model \cite{Pogosian:2006hg}, with the vector component being of greater importance and having a quite different form. Reducing the so-called wiggliness parameter in the unconnected segment model can to a certain extent boost the vector contribution relative to the scalar component, as can be seen in Fig. 1 of 
\cite{Pogosian:2006hg}, but the model cannot include velocity correlations (or the angular momentum of the decay products) which source the vector mode.

In this discussion of our polarization results, we additionally note that there is a similar shift in angular scale in the vector-dominated \BB\ spectrum. Our string simulations give a \BB\ peak at $\ell = 600$, with half of the peak value given at $\ell = 300$ and $1100$. On the other hand, the simpler model gives a peak at $\ell = 800$ with the half of the peak value given at $\ell = 400$ and $1500$. This may be important observationally because our results highlight a more significant difference in form between the cosmic string contribution and that from gravitational lensing.

The amplitude of the power spectrum also differs between the two calculation methods, with the unconnected segment model giving values for both peaks that are approximately 10 times greater than ours for a given $G\mu$ value. However, the normalization of the temperature power spectrum is also larger for that model and given constraints provided by current temperature data this effect is largely removed, with the factor of 10 becoming a mere doubling. 

Hence it would appear that the computationally rapid unconnected segment method does yield qualitatively correct polarization results but with quantitative differences relative to the results from our more complete simulations.


\section{Comparison with global defects}

Global defects result from the spontaneous breaking of global symmetries in the early universe, and share the important scaling property with cosmic strings. Unlike their local counterparts, global defects do not localize their energy into the defect cores and, for example, in the case of global strings, the strings themselves do not need to be resolved in the simulations. Hence, without the need for our string width control formalism (see Sec. \ref{sec:method}), temperature power spectra components from field theory simulations have previously been published for global textures \cite{Pen:1997ae, Durrer:1998rw, Bevis:2004wk, Bevis:2006mj}, global monopoles \cite{Pen:1997ae} and global strings \cite{Pen:1997ae}. Polarization spectra for these three global defect types have also been presented previously \cite{Seljak:1997ii} and we have applied our unequal-time correlation software and modified \CMBEASY\ code to the case of global textures to provide one means of checking our results against independent polarization calculations, and find excellent agreement.

\begin{figure}
\resizebox{\columnwidth}{!}{\includegraphics{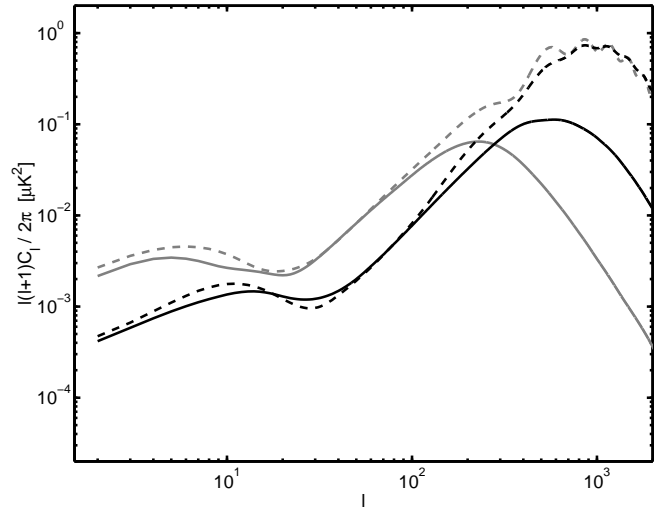}}
\caption{\label{fig:compGlobal}A comparison of the \EE\ (dashed) and \BB\ (solid) power spectra contributions from local cosmic strings (black) and global textures (gray). Each are calculated at $h=0.72$, $\Obhh=0.0214$, $\Ol=0.75$ and $\tau=0.1$, with the normalizations set to yield fractional contributions to the temperature power spectrum of $\fs=0.11$ for strings and $\fs=0.13$ for textures, which correspond to the $95\%$ upper bounds from \CMB\ data.}
\end{figure}

This also enables us to consider the differences in the polarization signals from local strings and global textures, both calculated at modern cosmological parameter values and shown in Fig. \ref{fig:compGlobal}. In the figure we have normalized the global texture contribution to match the $95\%$ upper bound found in \cite{Bevis:2004wk} using \CMB\ data and the above measurements of $h$ and $\Obhh$. For the temperature power spectrum (see \cite{Bevis:2006mj}), the key difference is that the textures yield a large bias toward low multipoles in comparison to local strings, giving a broad peak between $\ell = 20$ and $\ell = 300$ ($75\%$ of peak value). This trend can also be seen in both the \EE\ and \BB\ polarization spectra, with the later peaking at $\ell=200$ in the texture case, rather than at $\ell=600$ as it does for strings. 

This can be understood as follows. The decay of topological defects and the smoothing of the associated fields is limited by the causal horizon (or more precisely the distance that light could have traversed since inflation ended). In the global case there are large-scale gradients which constitute a form of potential energy. The dynamics of the system therefore rapidly removes the gradients, smoothing out the textures on scales just a little smaller the horizon and textures only contribute to energy and momenta for $k\tau\lesssim 10$ \cite{Bevis:2006mj,Durrer:1998rw}. In the local string case however, a gauge field is present which can accommodate the large-scale gradients such that they are effectively smoothed and contribute little energy. The strings themselves represent spatially extended potential energy contributions and the Hubble-damped dynamics tends to reduce the string length, but this process is less efficient so strings persist within the horizon. Hence, in contrast to the global case the Abelian Higgs fields contribute energy and momenta for all $k\tau\lesssim 100$ \cite{Bevis:2006mj}. Hence, the \CMB\ anisotropies sourced by the energy-momentum tensor of the local string model are more biased to smaller scales than for global textures, and global defects in general.


\section{Current and future \CMB\ measurements}

These first polarization results from string network simulations are very encouraging since although a finite \B-mode polarization has yet to be detected, there are a number projects in preparation that aim to measure \B-mode polarization with high precision. Current \CMB\ polarization data gives merely an upper bound on the \BB\ spectrum, with for example the full-sky \WMAP\ project providing a upper limit of $3\times10^{-2} \mu K^{2}$ derived from $2 \leq \ell \leq 12$. This is only a factor of 3 above the plotted tensor mode prediction for $r=0.36$, but about 30 times the mean value for $G\mu=0.7\times10^{-6}$ cosmic strings over this range. On smaller angular scales, the ground-based \DASI\ and balloon-borne \BOOMERANG\ detectors place limits on the \BB\ spectrum of a few $\mu K^{2}$ for multipoles $200 \lesssim \ell \lesssim 600$ \cite{Leitch:2004gd, Montroy:2005yx} while the ground-based CBI project similarly constrains multipoles $400 \lesssim \ell \lesssim 1700$ \cite{Sievers:2005gj}. These experiments were designed to detect to the \EE\ and \TE\ spectra, and were successful in this aim, but the detection of \B-mode polarization will require the next generation of sub-orbital \CMB\ experiments and, on the largest angular scales, the forthcoming Planck satellite mission.

Planck is scheduled for launch in 2008 and while its polarization sensitivity and angular resolution are significant improvements over \WMAP, it is not specifically designed to detect very small \B-mode signals. It will, however, provide a measurement of the \BB\ spectrum at large angular scales that will not be superseded until a subsequent \CMB\ satellite is launched, but this is more relevant for the detection of the primordial tensor mode than it is for cosmic strings. On smaller scales, Planck will struggle to detect the gravitational lensing signal, with very large uncertainties even after binning across large ranges in $\ell$. However, if strings contribute at the level plotted, they will be detected by Planck's \BB\ measurements, as what might be otherwise interpreted as a gravitational lensing signal that is 3 times larger than expected. Of course, they would probably also be significantly detected as a result of the precise \TT, \TE\ and \EE\ data that Planck will provide. With full-sky \TT\ data that is cosmic variance limited to multipoles possibly as high as $\ell=2500$ and the \EE\ power spectrum measured with $\sim10\%$ accuracy out to $\ell\gtrsim1000$, the cosmological parameters will be heavily constrained. Then, the \TT\ data should be sensitive to  smaller string contributions than the current data.

\begin{figure}
\resizebox{\columnwidth}{!}{\includegraphics{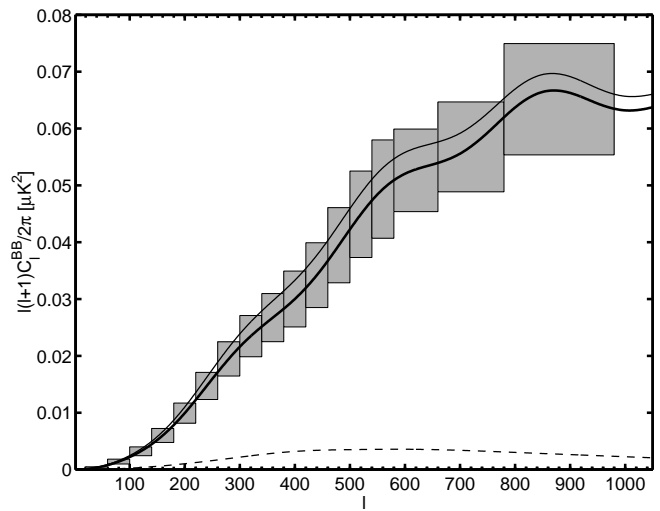}}
\caption{\label{fig:clover}Estimated uncertainties from the future \CLOVER\ project under the null hypothesis, that is that the true underlying model contains neither cosmic strings nor primordial tensor modes. The thick solid line indicates the underlying model, while the squares show the $\ell$-binning of the data and the 1$\sigma$ uncertainties estimated. The thin solid line shows the total power spectrum from an $\fs=0.0035$ model while the dashed thin shows the string component in that case.}
\end{figure}

However, planned sub-orbital projects have the potential to be more sensitive to the \B-mode polarization than the Planck satellite and, while they will not provide data for the very largest angular scales, it is the range $100 < \ell < 700$ that is most interesting for cosmic strings. These projects can deploy new technology more rapidly than a satelite mission and they can also choose the view fields to minimize foreground contamination. With the list of such projects including \cite{Oxley:2005dg, Taylor:2006jw}
: \EBEX\ (first flight 2008), \POLARBEAR\ (operational in 2008), \QUIET (deployment in early 2008), \CLOVER\ (operational in 2009), \POLARBEAR\ II (operational in 2010), and \SPIDER\ (first long-duration flight 2010); there should be a wealth of \B-mode data available in the relatively near future.

We will consider just one of these as an example, \CLOVER: a ground-based project that will measure the \B-mode polarization over the range $20<\ell<1000$ and is scheduled to become operational in 2009. The aim is to eliminate detector noise and foreground contamination to yield the precision required to detect primordial tensor modes even if $r$ is as low as $0.01$ \cite{Taylor:2006jw}. However, it will also characterize the high $\ell$ range and will be sensitive to string components that are very small in the \TT\ spectrum. 

Given the sensitivity, resolution, and sky coverage of a \CMB\ project, the likely uncertainties in the measured $C_{\ell}^{\mathrm{BB}}$ are approximately given by \cite{Kinney:1998md,Knox:1995dq,Kamionkowski:1996ks}:
\begin{equation}
\label{eqn:clover1}
	\sigma_{\ell}^{2} 
	= 
	\frac{2}{(2\ell+1) f_{\mathrm{sky}} \Delta\ell}
	\left( C_{\ell} + N_{\ell} \right)^{2}.
\end{equation}
The first term is due to (Gaussian) cosmic variance, while the second is the thermal noise:
\begin{equation}
	N_{\ell} 
	=
	\theta^{2}_{\mathrm{FWHM}}\sigma_{\mathrm{pixel}}^{2} \; \exp \left[ \frac{ \ell(\ell+1)
	\theta^{2}_{\mathrm{FWHM}}}{ 8 \ln 2} \right].
\end{equation}

The \CLOVER\ project is expected to cover a fraction $f_{\mathrm{sky}}=0.025$ of the sky which means that measurements at each $\ell$ are not independent, with correlations extending roughly $\Delta\ell=40$. The equation assumes a Gaussian beam profile, which we take to have $\theta_{\mathrm{FWHM}}=8'$ and then gives map pixels representing solid angle $\approx\theta_{\mathrm{FWHM}}^{2}$. We assume successful foreground removal and take the thermal pixel noise to be $\sigma_{\mathrm{pixel}}=0.65\,\mu K$, which corresponds roughly to that expected for \CLOVER\ after 2 years of observation. We then suppose that both primordial tensor modes and cosmic strings make a zero (or negligible) contribution --- the \emph{null hypothesis} --- and insert the corresponding \BB\ power spectrum into Eqn. \ref{eqn:clover1}, with the results shown in Fig. \ref{fig:clover}.

To make a crude estimate of the likely upper bound on the string contribution under the null hypothesis, we perform a one-dimensional likelihood analysis, varying $\fs$. We reduce the normalization $\As^{2}$ of the primordial scalar component, such that it is proportional to $(1-\fs)$ but keep all other cosmological parameters fixed. This is partially justifiable since, by the time of the \CLOVER\ two-year data release, the \TT, \TE\ and \EE\ data from the \CMB, combined with other independent data, will impose tight limits on them under the standard empirical models. Also the dependence of $\As^{2}$ upon $\fs$, which strictly could deviate from the chosen form for a fit over a range of multipoles, was found in \cite{Bevis:2007gh} to be approximately $(1-f_{10})$ for current data. We then find that the $95\%$ upper bound on the string contribution stemming from the \BB\ data will be $\fs<0.0035\pm0.0012$, with the central value being a factor of 30 less than the limits given by current data \cite{Bevis:2007gh}. Note that there is a statistical uncertainity in this result because the measured data will be spread around the underlying model according to the errors indicated in the figure, and different realizations will yield different upper bounds. The corresponding string tension limit is $G\mu<(0.12\pm0.02)\times10^{-6}$.

This is merely a first step towards a full investigation of possible future data, which would involve many different projects and a full multi-parameter analysis, but the approximate result here suggests that projects such as \CLOVER\ will be able to either detect cosmic strings from their \B-mode contribution or place stringent limits on cosmic string scenarios. 

Note that in contrast with \cite{Seljak:2006hi}, we have not investigated the cleaning of the gravitational lensing signal from the \CLOVER\ measurements, which in the standard case is partially achievable via the non-Gaussianity introduced by the lensing \cite{Hu:2001kj,Hirata:2002jy,Hirata:2003ka}. This is because the cosmic strings will introduce non-Gaussian signatures themselves and the non-Gaussianity from strings is not well characterized at present. It should also be noted that our results are not greatly changed by the inclusion of a small tensor contribution, since tensor modes contribute only to the largest scales, while supersymmetric hybrid inflation models [10], for example, do predict the negligibly small tensor mode assumed here.


\section{Conclusion}

We have presented the first \CMB\ polarization calculations to involve simulations of a (local) cosmic string \emph{network}, rather than using unconnected string segments with stochastic velocities. We have demonstrated, both here and in a sister paper \cite{Bevis:2006mj} in which we presented the method and temperature results, that field-theory simulations of local cosmic strings can be employed for this purpose with current computational facilities. Our results show that the computationally simpler unconnected segment method, which has been commonly used by other authors, gives results that are qualitively accurate. This had previously been confirmed \emph{only} for the temperature power spectrum. However, our results do show a greater bias to large angular scales. 

Importantly, we confirm the prediction of a large string contribution to the \B-mode polarization power spectrum, even for small contributions to the temperature anisotropies, with current upper bounds from the temperature data allowing a possible dominance of the \B-mode data by cosmic strings. Through simulating future data for the \CLOVER\ project we have shown that the likely \B-mode measurements appear sensitive to cosmic strings with fractional contributions to the temperature power spectrum at $\ell=10$ as low as $\fs<0.0035$, which is a factor of $30$ tighter than the current \CMB\ bounds. This corresponds to an upper bound on the string tension of $G\mu<0.12\times10^{-6}$.

This is very encouraging because a detection of cosmic strings would not only be interesting in itself, it would open a powerful window upon early universe physics. We have included here a comparison between the predictions for local U(1) cosmic strings and global defects, but different string models would of course also yield different results. For example, it would be interesting to perform \CMB\ calculations for semilocal string networks \cite{Achucarro:1999it,Hindmarsh:1991jq,Hindmarsh:1992yy,Urrestilla:2004eh,Dasgupta:2004dw,Achucarro:2005tu,Urrestilla:2007}, and also for simple models that yield Y-junctions between strings, such as the U(1)$\times$U(1) dual Abelian Higgs model \cite{Saffin:2005cs,Urrestilla:2007inprep} or the non-Abelian SU(2)/$\mathbb{Z}_{3}$ model of \cite{Hindmarsh:2006qn}. In string theory, composites of D- and F-strings are possible \cite{Copeland:2003bj}, with Y-junctions where they separate into their constituents. If differences between the \B-mode spectra of superstring-inspired models and traditional U(1) strings could be established, then the detection of one or other type of cosmic string might represent an exciting observational test for string theory.

Finally, we note that the \CMB\ power spectra are not the only means to constrain or detect cosmic strings. Strings are likely to produce significant non-Gaussianity in the \CMB\ and are not fully described by power spectra alone. In fact, cosmic strings would create discontinuities in \CMB\ maps of size $\Delta T/T \sim 13 G\mu$ \cite{Vilenkin:1994book}. These will only become directly noticeable once the pixel noise is of the same order as the discontinuities and the beam width of the microwave detector is sufficiently small for the inflation-induced \CMB\ temperature variations on sub-pixel scales (which contribute an effective noise) to be similarly reduced. However, a statistical approach may be more sensitive to the string non-Gaussianity since it can draw information from a large number of pixels and hence is an interesting avenue for future research.

\begin{acknowledgments}

We acknowledge support from \textsc{pparc} (NB, MH), the Swiss \textsc{nsf} (MK), Marie Curie Intra-European Fellowship MEIF-CT-2005-009628 and FPA2005-04823 (JU). For their hospitality during the final stages of this work, we thank the University of Geneva (MH, JU). The Abelian Higgs simulations were performed on \textsc{cosmos}, the UK National Cosmology Supercomputer, supported by \textsc{sgi}, Intel, \textsc{hefce} and \textsc{pparc}. We would like to thank Stuart Rankin and Victor Travieso of \textsc{cosmos}, as well as Anthony Challinor and Ruth Durrer for useful discussions. 

\end{acknowledgments}


\appendix

\section{Uncertainty estimation}

In this appendix we detail estimates of the uncertainties in our results, however, we also refer the reader to \cite{Bevis:2006mj} for a more in-depth discussion of both the method for calculating the power spectra and the limitations therein.

\begin{figure}
\resizebox{\columnwidth}{!}{\includegraphics{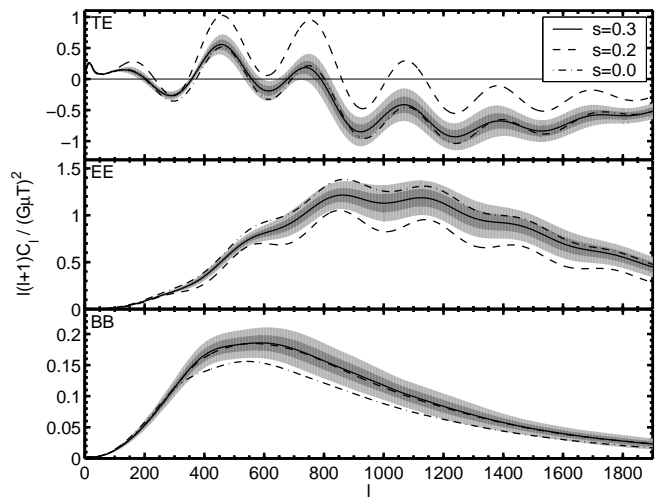}}
\caption{\label{fig:stat}The effect of $s$ on the results compared to the estimated uncertainties from realization-to-realization variations, with shaded areas showing the 1 and 2-$\sigma$ regions at each multipole. Note that correlations extend over large multipole ranges, that there are statistical uncertainties for each $s$, and that $T$ is the mean CMB temperature.}
\end{figure}

An obvious source of uncertainty in our results stems from the fact that the initial conditions for the string simulations are randomly generated. Therefore the statistical results taken from the simulations and used in the \CMBEASY\ calculations will differ from realization to realization. We averaged the simulation results over 5 realizations (in each of the matter and radiation eras), which amounted to 625 \CPU-days of processing time on the UK National Cosmology Supercomputer \cite{cosmos-website}. Hence we may estimate the variance in our power spectra by additionally applying the modified \CMBEASY\ code upon the results from individual realizations, giving the results shown in Fig. \ref{fig:stat}. In the case of the \TE\ power spectrum, the estimated standard deviation in the mean over the 5 realizations is about $20\%$ for $\ell\gtrsim1000$ and is less than $10\%$ for $\ell<100$ but with larger relative values in the intermediate regime where zero-crossings are common. In the \EE\ spectrum, we find relative values between $4$ and $11\%$ over the range $2<\ell<3000$ while in the \BB\ spectrum we find between $5$ and $10\%$ for $\ell\lesssim1000$, before this value  increases roughly linearly with $\ell$ to around $30\%$ at $\ell=3000$. 

As discussed in Sec. \ref{sec:method}, the computational limitations and the rapid growth of the horizon size relative to the string width meant that the true Abelian Higgs dynamics had to be approximated such that the string width grew in physical units in proportion to $a^{1-s}$. The closest to the true $s=1$ case that was achievable in the matter era was $s=0.3$ and hence we investigated the effect of $s$ upon our results in order to estimate the corresponding systematic error. It can be seen in Fig. \ref{fig:stat}, that in the \TE\ case there is a large change from the $s=0.3$ and $s=0.2$ cases and yet the $s=0$ case lies very close to that for $s=0.3$. Further it should be noted that there are statistical uncertainties on all three lines and the results are consistent with the seen deviations between the three cases being largely due to the natural realization-to-realization variations rather than to any sensitive $s$ dependence. 

More evidence in support of this conclusion comes from the behavior of the strings in the radiation era, which can be studied under $s=1$ for a meaningful period of time. The results for both the string length density and the equal-time correlation functions of the energy momentum tensor are shown in \cite{Bevis:2006mj} to deviate from the $s=0$ case by an amount of the the same order as the statistical uncertainties, with non-resolvable differences between the $s=0$, $s=0.2$ and $s=0.3$ cases. Hence any systematic errors in the \TE\ power spectrum results due to the use of $s<1$ are likely to be of the same order as the measured statistical uncertainties, and this is also true for the \EE\ and \BB\ spectra.

A further source of uncertainty stems from the possibility that the approximate scaling behavior seen is not actually the true behavior in the large $\tau$ limit. While in principle this could be because the string decay products included in the simulations are those inherent to the Abelian Higgs model rather than gravitational radiation which may in fact be more important. Unfortunately a detailed knowledge of the string dynamics on scales much smaller than the horizon is required before the nature of the primary decay products of cosmic strings are understood and there is no numerical estimate that is readily achievable for this factor. However, even ignoring the possibility of a important back reaction from gravitational radiation, the seen scaling behavior may still be in error. 

Under the expected scaling behavior, the mean (comoving) length of string $L_{\mathrm{H}}$ per (comoving) horizon volume $V_{\mathrm{H}}$ should be proportional to $\tau$, while $V_{\mathrm{H}}$ is proportional to $\tau^{3}$ and hence:
\begin{equation}
\xi = \sqrt{\frac{ V_{\mathrm{H}} }{ L_{\mathrm{H}} }} \propto \tau.
\end{equation}
However, this was not the scaling behavior seen and instead, $\xi$ was found to vary as:
\begin{equation}
\xi \propto (\tau - \tauOS),
\end{equation}
in the scaling era \cite{Bevis:2006mj}. While this is not a problem, since at the late times required for \CMB\ calculations the offset $\tauOS$ is negligible, it does mean that every $\tau$ used in the calculation of the unequal-time scaling functions must be replaced by $\tau - \tauOS$ in order for them to be representative of the late time behavior. If there is any curvature in the supposedly linear region, then the measured $\tauOS$ is be a function of the $\tau$ range over which a linear fit is applied. For the final results presented herein we fit over the range $64\VEV^{-1} \leq \tau \leq 128\VEV^{-1}$, where $\VEV$ is the energy scale in the Abelian Higgs model (see \cite{Bevis:2006mj}). By varying the upper limit of the fitting range a ball-park estimate of the effect may be obtained, with the results shown in Table \ref{tab:errors} for $\ell=300$, which is roughly where the contribution to the \BB\ spectrum from strings is most significant. 

\begin{table}
\begin{ruledtabular}
\begin{tabular}{lcccc}
Source or change                                       & \TE\     & \EE\     & \BB\ \\
\hline
\\
Statistical variations                                 & $\pm13$ & $\pm4.6$ & $\pm5.5$ \\
\\
$\xi$ fit $\tau_{\mathrm{max}} = 128 \rightarrow 112$  & $+6.3$   & $-3.7$   & $-6.6$ \\
$\xi$ fit $\tau_{\mathrm{max}} = 128 \rightarrow \;96$ & $+13.5$   & $-3.7$   & $-6.0$ \\
\\
$\Rmax \approx 1.8 \rightarrow 1.6$         & $-48$  & $-0.17$   & $+2.7$ \\
$\Rmax \approx 1.8 \rightarrow 1.5$         & $-63$  & $-2.3$    & $+2.3$ \\
$\Rmax \approx 1.8 \rightarrow 1.4$         & $-44$  & $-5.5$    & $+4.5$ \\
\\
Matrix size $M=512 \rightarrow 256$         & $+8.5$ & $+0.15$  & $-1.4$ \\
\\
Radiation data replaced by matter data      & $+25$    & $-13$  & $-17$ \\
\\
\end{tabular}
\end{ruledtabular}
\caption{\label{tab:errors}Investigations into the uncertainties in the CMB power spectra contributions from cosmic strings at $\ell=300$. The estimated uncertainty, or the responses to changes in the calculation procedure, are shown as percentages.}
\end{table}

An additional systematic uncertainty arises from the limited range of time ratios at which the unequal time correlation functions can be calculated. As explained in \cite{Bevis:2006mj}, the data is taken from the simulations over the period in conformal time: $64\VEV \leq \tau \leq 160\VEV$. This is the time period after the strings have formed and are scaling to well within the statistical uncertainties but before the simulation becomes noticeably aware of the periodic boundary conditions. It gives a maximum time ratio of $160/64=2.5$, however, the offset scaling law and the incorporation of $\tauOS$ decreases this value to around $1.8$. Hence, we cannot study the scaling regime over large ranges of time. Fortunately, the unequal-time correlation functions decay for large and small time ratios (see \cite{Bevis:2006mj}) and hence the most important data can be taken from the simulations that were possible and we took the correlation functions to be zero for more extreme time ratios than we were able to study.

The error instilled by this can be investigated by further zeroing known regions of the correlation functions, above a time ratio $\Rmax$ and below a time ratio $\Rmax^{-1}$. We explored the change in the power spectra for truncations with $\Rmax=1.6$, 1.5 and 1.4, which led to the changes shown in the table. The \TE\ results are very sensitive to $\Rmax$ and a further investigation may be required before a comparison of these results against, for example, Planck data can be reliably performed. A similar situation exists for the \EE\ data over certain $\ell$ ranges: $\ell\sim20$ and $\ell\gtrsim1000$, however, again it will be some time before data exists that will necessitate \TE\ and \EE\ calculations for strings of greater precision. Fortunately, the more important \BB\ spectra are robust, with the present results appearing more than adequate for comparisons to \BB\ data from, for example, \CLOVER. 

A further source of uncertainty arises as a result of possible numerical errors during the eigenvector decomposition phase. The simulations output data for a particular timestep $\tau$ against a reference time $\tau'$ for a series of $k$ values and hence output $\FT{C}(k\tau,\tau/\tau')$ at discrete values of its two inputs. However the eigenvector decomposition requires the data to be available at discrete $k\tau$ and $k\tau'$ to form a matrix of size $M\times M$. There is hence an incorporation of numerical errors during the required interpolation procedure that is more important for smaller $M$. Further, the resulting eigenvectors have a $k\tau$ resolution that is dependent upon $M$ (if linear $k\tau$ spacing is used over a given $k\tau$ range) and interpolation is carried out to provide data for the modified $\CMBEASY$ code at a given timestep $\tau$ when solving a given $k$ mode. We hence explored a reduction in $M$ from the value of 512 that is used for our primary results, giving the changes shown in the table at $\ell=300$ and changes less than about $10\%$ in the \EE\ and \BB\ spectra for all $\ell<1000$, and in the \TE\ spectra for $\ell<100$ (before the oscillating regime begins). Note that larger values of $M$ necessitate a larger number of eigencontribution terms to be included inorder to yeild convergence of the power spectra sums to well below the other error estimates.

Finally, we investigate the possible errors in our modeling of the radiation-matter transition, at which scaling is broken and we interpolate between in the eigenvectors (and eigenvalues) calculated in each era (see \cite{Bevis:2006mj}). An \emph{overestimate} of the effect can be made by replacing the radiation era data with that from the matter era, yielding the results shown in the table for $\ell=300$. However, the change is much less important at large angular scales where the impact of the radiation era is smaller. 


\bibliography{CMBpolarizationFromStrings}

\end{document}